\documentclass[sigconf]{acmart}

\usepackage{amsmath}
\usepackage{booktabs}
\usepackage{subcaption}
\usepackage{algorithm}
\usepackage{algorithmic}
\usepackage{xcolor}
\usepackage{balance}

\microtypesetup{protrusion=false}

\newcommand{\rev}[1]{#1}
\usepackage{enumitem}
\setlist[itemize]{leftmargin=*, topsep=2pt, itemsep=1pt, parsep=0pt, partopsep=0pt}

\newtheorem{definition}{Definition}[section]
\newtheorem{problem}{Problem}[section]
\newtheorem{theorem}{Theorem}[section]

\copyrightyear{2026}
\acmYear{2026}
\setcopyright{cc}
\setcctype{by}
\acmConference[CIKM '26]{Proceedings of the 35th ACM International Conference on Information and Knowledge Management}{November 07--11, 2026}{Rome, Italy}
\acmBooktitle{Proceedings of the 35th ACM International Conference on Information and Knowledge Management (CIKM '26), November 07--11, 2026, Rome, Italy}
\acmDOI{10.1145/3799682.3840684}
\acmISBN{979-8-4007-2539-5/2026/11}

\begin{document}

\title{\textsc{PACIFIER}: Pacing Opinion Depolarization via a Unified Graph Learning Framework}

\author{Mingkai Liao}
\affiliation{%
  \institution{Shenzhen University}
  \city{Shenzhen}
  \country{China}
}
\email{13542830854@163.com}

\author{Zijun Zhang}
\affiliation{%
  \institution{Hubei University of Police}
  \city{Wuhan}
  \country{China}
}
\email{zhangzijunhbpa@foxmail.com}

\author{Mingyang Zhou}
\authornote{Corresponding author.}
\affiliation{%
  \institution{Shenzhen University}
  \city{Shenzhen}
  \country{China}
}
\email{zmy@szu.edu.cn}

\renewcommand{\shortauthors}{Mingkai Liao, Zijun Zhang, and Mingyang Zhou}

\begin{abstract}
Online social networks often form opposing echo chambers that reinforce opinion
polarization. Under the Friedkin--Johnsen (FJ) model,
\textsc{ModerateInternal} (MI) and \textsc{ModerateExpressed} (ME) reduce
polarization by neutralizing selected users' internal or expressed opinions,
but existing solutions are largely model-specific and learning-based
depolarization remains underexplored. We study FJ-based moderation as
\rev{feedback-free autoregressive sequential planning} and propose PACIFIER, a
unified graph-learning framework that constructs ordered intervention sequences
from the initial graph--opinion instance without using recomputed
post-intervention expressed opinions as intermediate feedback. PACIFIER combines
history-aware node representations with Greedy and RL variants for immediate
and long-horizon action scoring, and supports MI, ME, continuous opinions,
cost-aware moderation, and node removal. We evaluate intervention trajectories
by Accumulated Normalized Polarization (ANP). Trained only on synthetic graphs
with fewer than 50 nodes, PACIFIER transfers to 15 real-world Twitter networks
with up to 155{,}599 nodes. It improves over the strongest non-PACIFIER baseline
by up to 35.3\%, while PACIFIER-RL improves over PACIFIER-Greedy by up to
37.69\%. Compared with oracle-style Greedy, PACIFIER retains 98.94\% and
97.39\% near-oracle quality on \textsc{MI} and \textsc{ME}, respectively, and
is about $600\times$ faster at the 450-node range. The source code is available
at \url{https://github.com/RickyYY-SZU/PACIFIER}.
\end{abstract}



\maketitle

\section{Introduction}
\label{sec:introduction}

Online social networks are central to public discussion and information
diffusion, but they can also reinforce opinion polarization through
like-minded interaction and algorithmically mediated exposure
\cite{sunstein2002law,garrett2009echo,bakshy2015exposure}. Empirical studies
likewise observe segregated political interaction patterns in retweet,
endorsement, and follow networks
\cite{conover2011political,adamic2005political,garimella2018quantifying}. This
motivates a central algorithmic question: under a limited intervention budget,
which users or actions should be selected to reduce polarization on large
social networks?

The Friedkin--Johnsen (FJ) model~\cite{friedkin1990social} distinguishes
persistent internal opinions from socially influenced expressed opinions.
Matakos et al.~\cite{matakos2017measuring} define an FJ-equilibrium polarization
index and two NP-hard benchmarks: \textsc{ModerateInternal} (MI), which
neutralizes selected users' internal opinions, and \textsc{ModerateExpressed}
(ME), which fixes selected users' expressed opinions to neutrality. Related FJ
interventions study polarization, conflict, link modification, and online
settings~\cite{musco2018minimizing,chen2018quantifying,zhu2022nearly,garimella2017reducing,zhu2021minimizing,racz2023towards,xu2023minimizing,cinus2026online},
but existing methods are often tied to specific objectives, intervention
primitives, or information regimes. Many MI/ME methods also repeatedly
recompute settled expressed opinions after partial interventions for candidate
evaluation or policy input, which is costly on large graphs.

These limitations motivate \emph{\rev{feedback-free autoregressive sequential
planning}}: a planner observes the initial graph--opinion instance,
\rev{selects one intervention at a time while updating only deterministic
bookkeeping information}, and constructs the sequence without recomputed
post-intervention expressed opinions as intermediate feedback. We evaluate the
trajectory by \emph{Accumulated Normalized Polarization} (ANP), which rewards
early and persistent polarization reduction. Fixed-topology MI/ME also create a
state-aliasing challenge because adjacency does not record which nodes have
already been intervened on, motivating explicit initial-opinion anchors and
selection-history features.

We propose \textbf{PACIFIER}, a unified graph-learning framework that preserves
canonical MI/ME semantics while lifting endpoint set selection to
\rev{feedback-free autoregressive sequential planning}. A shared encoder--decoder
uses history-aware node features, initial-opinion anchors, selected-node marks,
feasibility masks, scale-normalized auxiliary signals, and a virtual super-node.
\textbf{PACIFIER-Greedy} learns immediate action scores, whereas
\textbf{PACIFIER-RL} learns long-horizon action values; the same interface
supports MI, ME, cost-ME, continuous-ME, and topology-modifying node removal.

PACIFIER is trained only on synthetic graphs with fewer than 50 nodes and
evaluated on 15 real-world Twitter follow and retweet networks with up to
155{,}599 nodes. \rev{Although trained on synthetic two-camp graphs, PACIFIER
directly generalizes to previously unseen real-world networks with a similar
two-echo-chamber polarization structure, without target-specific retraining.}
PACIFIER remains competitive with model-aware MI solvers and consistently
improves over non-PACIFIER baselines in ME and the extended settings; on
real-world ME and node removal, average ANP improves by 21.1\% and 19.2\% over
PageRank, respectively. Full-information comparisons further show that
\rev{feedback-free} PACIFIER is within 0.21\% of its full-information counterpart
while avoiding repeated equilibrium recomputation. \rev{These results also help
clarify the scope of PACIFIER: its core assumption is that polarization is
driven by a two-echo-chamber structure, rather than by the linearity of the FJ
model. We use FJ dynamics for direct comparison with established depolarization
problems and baselines; for other opinion dynamics with the same structural
source of polarization, including nonlinear biased-assimilation
models~\cite{dandekar2013biased}, PACIFIER can be adapted by replacing the
transition dynamics and the corresponding reward/evaluation computation.}

Our contributions are fourfold. First, we reformulate FJ-based MI and ME as
\rev{feedback-free autoregressive sequential planning} evaluated by ANP. Second,
we introduce a shared PACIFIER node-action interface for Greedy and RL policies
across multiple intervention settings. Third, we address state aliasing in
fixed-topology continuous-state intervention with history-aware anchors, marks,
masks, and auxiliary graph signals. Fourth, experiments on synthetic and
real-world polarized networks demonstrate strong cross-scale performance and a
favorable \rev{feedback-free} quality--efficiency trade-off.

\section{Related Work}
\label{sec:related}

We organize related work along two threads: FJ-based polarization intervention,
which defines the moderation semantics used in PACIFIER, and graph-learning
based network intervention, which motivates history-aware sequential policies
and small-to-large transfer.

\noindent\textbf{Opinion-dynamics-based moderation.}
The Friedkin--Johnsen (FJ) model is a standard framework for opinion formation
in social networks~\cite{friedkin1990social}. By distinguishing persistent
internal opinions from socially influenced expressed opinions, it naturally
supports interventions on beliefs, public expressions, or network structure.
A broad line of work optimizes polarization, disagreement, controversy,
conflict, consensus, or related equilibrium quantities under opinion-dynamics
models, including connecting opposing views~\cite{garimella2017reducing},
minimizing conflict risk~\cite{chen2018quantifying,zhu2022nearly},
minimizing polarization--disagreement objectives~\cite{musco2018minimizing},
link recommendation~\cite{zhu2021minimizing}, network perturbation toward
consensus~\cite{racz2023towards}, leader--follower intervention
\cite{xu2023minimizing}, opinion optimization in directed networks
\cite{sun2023opinion}, and online intervention with unknown internal opinions
\cite{cinus2026online}. These methods demonstrate the effectiveness of
carefully chosen interventions, but are typically tied to specific objectives or
intervention primitives.

The most directly related formulation is due to Matakos et
al.~\cite{matakos2017measuring}, who define polarization using the FJ
equilibrium expressed-opinion vector and introduce two canonical moderation
problems: \textsc{ModerateInternal} (MI), which neutralizes selected users'
internal opinions, and \textsc{ModerateExpressed} (ME), which fixes selected
users' expressed opinions to neutrality. PACIFIER preserves these MI/ME
semantics, but evaluates ordered intervention sequences by ANP, supports
\rev{feedback-free autoregressive planning} without intermediate equilibrium feedback, and uses a shared
node-action interface that extends to cost-ME, continuous-ME, and node removal.
Learning-based depolarization remains limited: Mylonas and
Spyropoulos~\cite{mylonas2026opinion} train a GNN to approximate
immediate gains for ME problem. We include an
aligned full-information extension, \textbf{GNN-GreedyExt}, as a baseline, but
PACIFIER instead learns an ANP-aligned sequential policy with explicit
intervention-history encoding.

\noindent\textbf{Graph-learning-based intervention and transfer.}
Graph intervention tasks differ in how much action history is visible to a
message-passing encoder. Topology-modifying interventions, such as optimal
percolation, network dismantling, and edge addition or rewiring
\cite{morone2015influence,braunstein2016network,ghosh2006growing,
musco2018minimizing,zhu2021minimizing,racz2023towards}, directly change the
graph, so the residual topology records part of the intervention history.
Fixed-topology diffusion tasks, such as influence maximization
\cite{kempe2003maximizing}, often expose seed or activation states explicitly;
learning methods such as structure2vec-based RL, GCOMB, and DeepIM exploit
this setting for graph combinatorial optimization
\cite{dai2017learning,manchanda2020gcomb,ling2023deepim}.

PACIFIER targets a more delicate regime: fixed-topology continuous-state
equilibrium intervention. In MI and ME, actions modify node-level opinion
attributes or equilibrium constraints while the adjacency matrix may remain
unchanged, so topology alone does not reveal which nodes have already been
intervened on. This creates a state-observability challenge that does not arise
in the same form in topology-changing or cascade-based tasks. PACIFIER addresses
it with initial-opinion anchors, selected-node marks, feasibility masks, and
scale-normalized polarization features.

Cross-scale generalization is also central to PACIFIER. FINDER
\cite{fan2020finding} and GCOMB~\cite{manchanda2020gcomb} show that learned
graph policies can train on small synthetic graphs and transfer to larger real
networks when the training distribution, state representation, and action
semantics remain stable. PACIFIER brings this small-to-large transfer idea to
FJ-based polarization moderation through two-camp echo-chamber training,
history-aware state representation, normalized auxiliary signals, and
ANP-aligned rewards. Compared with analytical MI/ME solvers, GNN-assisted
greedy depolarization, and generic graph intervention learning, PACIFIER learns
a reusable \rev{feedback-free autoregressive policy} for topology-preserving continuous-state
moderation.

\section{Problem Formulation}
\label{sec:formulation}

This section formalizes the polarization moderation problem studied in
PACIFIER. We first define the Friedkin--Johnsen (FJ) opinion-dynamics
model~\cite{friedkin1990social} and the polarization index. We then introduce
the canonical \textsc{ModerateInternal} (MI) and \textsc{ModerateExpressed}
(ME) moderation tasks, lift them from endpoint set selection to ordered
intervention-sequence planning, and define Accumulated Normalized Polarization
(ANP) as the trajectory-level evaluation metric. Finally, we specify the
\rev{feedback-free} information constraint and the extended settings considered in this
paper.

Let $G=(V,E)$ be an undirected graph with $n=|V|$ nodes.  Let $\mathbf{s}^{(0)}\in[-1,1]^n$ denote the
initial internal-opinion vector, where $+1$ and $-1$ represent two opposing
extreme stances and $0$ represents a neutral stance.

\subsection{FJ Opinion Model and Polarization}
\label{subsec:fj_model}

Let $A\in\mathbb{R}^{n\times n}$ be the weighted adjacency matrix and
$D=\operatorname{diag}(d_1,\ldots,d_n)$ with
$d_i=\sum_{j\in N(i)}w_{ij}$. The weighted graph Laplacian is
$L_G=D-A$. Each node has a positive self-weight $\lambda_i>0$, collected in
$\Lambda=\operatorname{diag}(\lambda_1,\ldots,\lambda_n)$. In the canonical
unit self-weight setting, $\Lambda=I$.

In the FJ model, each node has a fixed internal opinion $s_i$ and an
expressed opinion $z_i$. The iterative update is
\begin{equation}
\label{eq:fj_update_general}
z_i^{(\tau+1)}=
\frac{\lambda_i s_i+\sum_{j\in N(i)}w_{ij}z_j^{(\tau)}}
{\lambda_i+\sum_{j\in N(i)}w_{ij}} .
\end{equation}
At convergence,
\begin{equation}
\label{eq:fj_equilibrium_general}
(L_G+\Lambda)\mathbf{z}^{(0)}=\Lambda\mathbf{s}^{(0)},\qquad
\mathbf{z}^{(0)}=(L_G+\Lambda)^{-1}\Lambda\mathbf{s}^{(0)} .
\end{equation}
Since $\Lambda$ is positive diagonal and $L_G$ is positive semidefinite,
$L_G+\Lambda$ is positive definite, so the FJ equilibrium is unique. Accordingly,
for any fixed graph--opinion state, FJ equilibrium evaluation is deterministic
and carries no simulation variance~\cite{matakos2017measuring}. Under
$\Lambda=I$, we write
$Q_{\mathrm{FJ}}=(L_G+I)^{-1}$ and
$\mathbf{z}^{(0)}=Q_{\mathrm{FJ}}\mathbf{s}^{(0)}$.

Following Matakos et al.~\cite{matakos2017measuring}, polarization is measured
by the squared distance of expressed opinions from neutrality:
\begin{equation}
\label{eq:polarization_index}
\pi_G(\mathbf{z})=\frac{1}{|V|}\|\mathbf{z}\|_2^2 .
\end{equation}
Lower values indicate greater population-level neutrality.

\begin{figure}[!tbp]
\centering

\begin{subfigure}{\linewidth}
    \centering
    \includegraphics[width=0.98\linewidth]{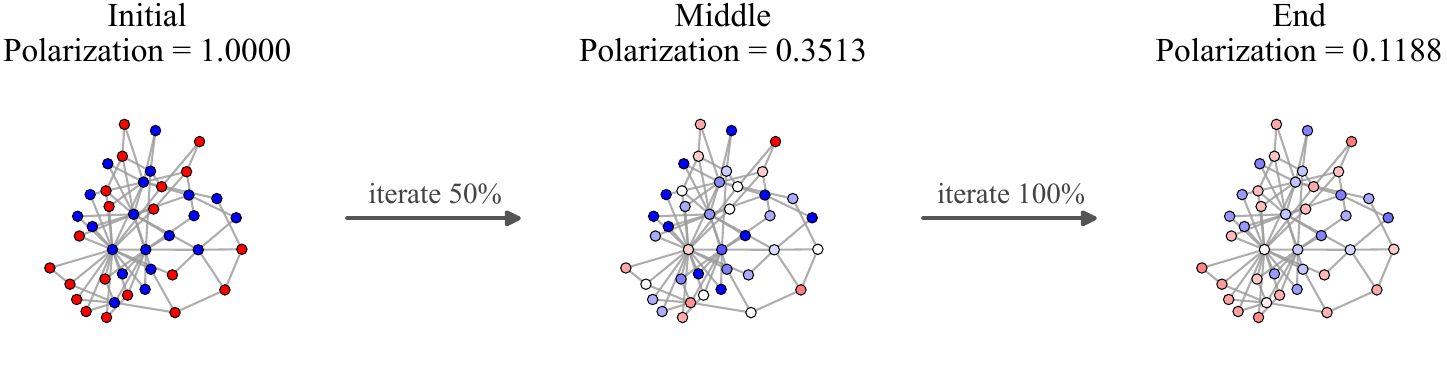}
    \caption{Randomly mixed opposing opinions.}
    \label{fig:fj_iterating_mixed}
\end{subfigure}

\vspace{0.2em}

\begin{subfigure}{\linewidth}
    \centering
    \includegraphics[width=0.98\linewidth]{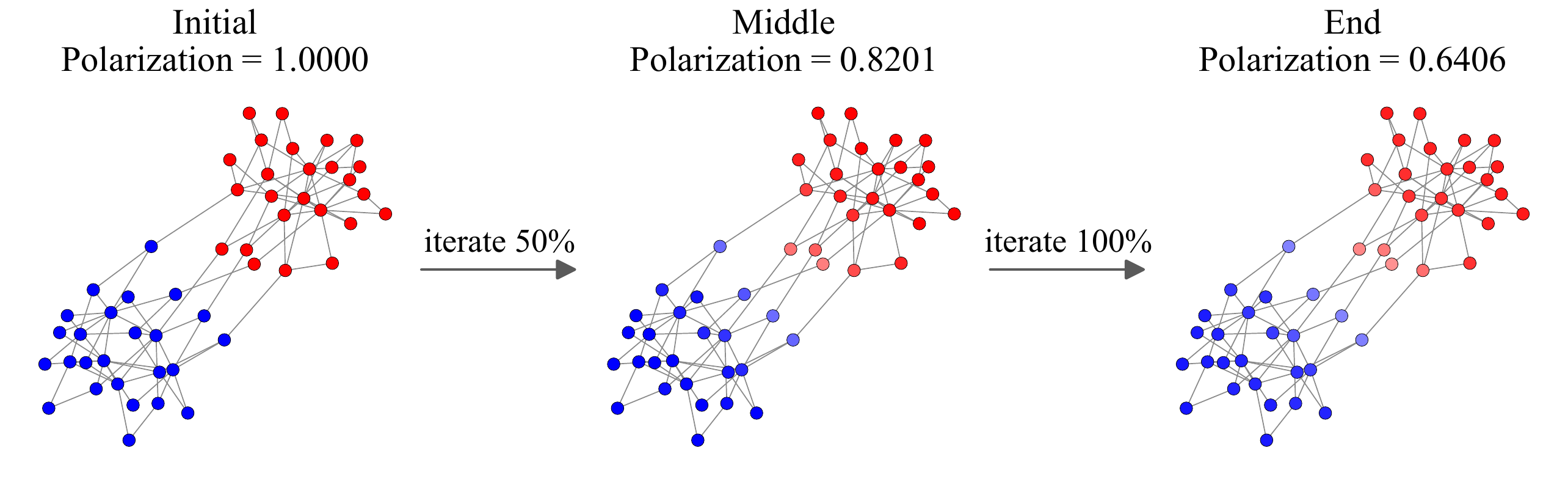}
    \caption{Camp-aligned opposing opinions on two echo chambers.}
    \label{fig:fj_iterating_aligned}
\end{subfigure}

\caption{FJ opinion iteration under two polarization regimes.}
\label{fig:fj_iterating_examples}
\vspace{-0.5em}
\end{figure}

FJ-based polarization intervention commonly uses two-camp echo-chamber
instances as polarized initial conditions. As illustrated in
Fig.~\ref{fig:fj_iterating_examples}, social averaging smooths randomly mixed
opposing opinions, but preserves substantial polarization when opinions align
with weakly connected camps. This setting is consistent with Matakos et
al.~\cite{matakos2017measuring} and R\'acz and
Rigobon~\cite{racz2023towards}, which study polarization intervention under
opposing camps separated by sparse cross-camp exposure. It is also consistent
with the Twitter controversy graphs of Garimella et
al.~\cite{garimella2018quantifying}, which provide the real-world polarized
networks used in our experiments. We formalize this initialization as follows.

\begin{definition}[Two-camp echo-chamber initialization]
\label{def:two_echo_chamber}
A graph--opinion instance $(G,\mathbf{s})$ has a two-camp echo-chamber
initialization if $V$ is partitioned into camps $V^+$ and $V^-$ with dominant
within-camp connectivity, sparse but nonempty cross-camp connectivity, and
camp-aligned internal opinions, i.e., $s_i=+1$ for $i\in V^+$ and $s_i=-1$
for $i\in V^-$. In continuous-opinion extensions, the sign of $s_i$ remains
camp-aligned while its magnitude may vary in $[-1,1]$.
\end{definition}

\subsection{Polarization Moderation and ANP Evaluation}
\label{subsec:moderation_anp}

We next define the moderation mechanisms and the trajectory-level evaluation
metric. The canonical MI and ME tasks specify the intervention semantics, while
ordered sequences, residual polarization, and ANP specify how an intervention
plan is evaluated over time.

\subsubsection{Canonical MI/ME Moderation}
\label{subsec:canonical_moderation}

The canonical MI and ME mechanisms are fixed-topology: they keep
$G=(V,E,w)$ fixed and modify only node-level opinion variables or equilibrium
constraints. After an intervention prefix $\sigma_t$, the settled expressed
opinion is written as
$\mathbf{z}_{\mathcal{M}}^{(t)}
=\operatorname{Eq}_{\mathcal{M}}(G,\mathbf{s}^{(0)},\sigma_t)$.

\begin{problem}[Moderating internal opinions]
\label{prob:mi}
\textsc{ModerateInternal}. Given $G=(V,E,w)$, initial internal opinions $\mathbf{s}^{(0)}$, and budget
$k$, find $T_s\subseteq V$ with $|T_s|=k$ such that polarization is minimized
after setting $s_i=0$ for all $i\in T_s$:
\begin{equation}
\label{eq:mi_objective}
\min_{T_s\subseteq V,\ |T_s|=k}
\pi_G\!\left(\mathbf{z}_{\textsc{MI}}(T_s)\right).
\end{equation}
\end{problem}

\begin{problem}[Moderating expressed opinions]
\label{prob:me}
\textsc{ModerateExpressed}. Given $G=(V,E,w)$, initial internal opinions $\mathbf{s}^{(0)}$, and budget
$k$, find $T_z\subseteq V$ with $|T_z|=k$ such that polarization is minimized
after fixing $z_i=0$ for all $i\in T_z$ and recomputing the FJ equilibrium:
\begin{equation}
\label{eq:me_objective}
\min_{T_z\subseteq V,\ |T_z|=k}
\pi_G\!\left(\mathbf{z}_{\textsc{ME}}(T_z)\right).
\end{equation}
\end{problem}

Problems~\ref{prob:mi} and~\ref{prob:me} specify the endpoint semantics of MI
and ME. PACIFIER keeps these semantics but evaluates an ordered intervention
trajectory rather than only the final selected set.

\begin{theorem}[\cite{matakos2017measuring}]
\label{thm:mi_me_np_hard}
The canonical endpoint versions of \textsc{ModerateInternal} and
\textsc{ModerateExpressed} are NP-hard.
\end{theorem}

\subsubsection{Ordered Sequences and Residual Polarization}
\label{subsec:rps}

Let $\mathcal{P}_k(V)$ denote the set of all ordered length-$k$ node sequences
without repeated nodes. For
$\sigma=(v_1,\ldots,v_k)\in\mathcal{P}_k(V)$, let
$\sigma_t=(v_1,\ldots,v_t)$ and $T_t=\{v_1,\ldots,v_t\}$ be its prefix and
selected-node set. We use $\sigma_0=\emptyset$ and $T_0=\emptyset$.

For \textsc{MI}, selected nodes have their internal opinions neutralized,
i.e., $s_i^{(t)}=0$ if $i\in T_t$ and $s_i^{(t)}=s_i$ otherwise. The
equilibrium is
\begin{equation}
\label{eq:mi_prefix_equilibrium}
\mathbf{z}^{(t)}_{\textsc{MI}}
=
(L_G+\Lambda)^{-1}\Lambda\mathbf{s}^{(t)} .
\end{equation}

For \textsc{ME}, selected nodes are fixed to neutral expressed opinions:
$\mathbf{z}^{(t)}_{T_t,\textsc{ME}}=\mathbf{0}$. Let
$F_t=V\setminus T_t$ and $H=L_G+\Lambda$. The equilibrium is
\begin{equation}
\label{eq:me_prefix_equilibrium}
\mathbf{z}^{(t)}_{F_t,\textsc{ME}}
=
(H_{F_tF_t})^{-1}\Lambda_{F_tF_t}\mathbf{s}_{F_t},
\qquad
\mathbf{z}^{(t)}_{T_t,\textsc{ME}}=\mathbf{0}.
\end{equation}

For any mechanism $\mathcal{M}$, the Residual Polarization Score (RPS) after
prefix $\sigma_t$ is
\begin{equation}
\label{eq:rps}
RPS_{\mathcal{M}}(\sigma_t)
=
\pi_{G_t}\!\left(\mathbf{z}^{(t)}_{\mathcal{M}}\right),
\end{equation}
where $G_t=G$ for MI and ME. For topology-modifying extensions, $G_t$ is the
current residual graph and the normalization in Eq.~\eqref{eq:polarization_index}
uses the current node set.

\subsubsection{Accumulated Normalized Polarization}
\label{subsec:anp}

ANP evaluates the area under the normalized residual-polarization curve. It
rewards sequences that reduce polarization early and keep it low. Computing
intermediate equilibria for ANP evaluation does not violate \rev{feedback-free planning},
because the restriction applies to sequence construction, not post-hoc
evaluation.

For $\sigma=(v_1,\ldots,v_k)$, ANP is
\begin{equation}
\label{eq:anp_metric}
ANP_{\mathcal{M}}(\sigma)
=
\frac{1}{k}\sum_{t=1}^{k}
\frac{RPS_{\mathcal{M}}(\sigma_t)}
{RPS_{\mathcal{M}}(\emptyset)+\varepsilon},
\end{equation}
where $\varepsilon>0$ prevents division by zero. A lower ANP indicates a better
moderation trajectory.

In cost-aware moderation, each node has cost $c(v)\ge 0$. Let
$\bar{c}=|V|^{-1}\sum_{v\in V}c(v)$ and
$\alpha(v)=c(v)/(\bar{c}+\varepsilon_c)$. The cost-aware objective is
\begin{equation}
\label{eq:cost_anp}
ANP_{\mathcal{M}}^{cost}(\sigma)
=
\frac{1}{k}\sum_{t=1}^{k}\alpha(v_t)
\frac{RPS_{\mathcal{M}}(\sigma_t)}
{RPS_{\mathcal{M}}(\emptyset)+\varepsilon}.
\end{equation}
When all costs are identical, Eq.~\eqref{eq:cost_anp} reduces to
Eq.~\eqref{eq:anp_metric}.

\subsection{\rev{Feedback-Free Autoregressive Sequential Planning} and Extended Settings}
\label{subsec:one_shot_and_settings}

We finally specify the information constraint under which intervention
sequences are constructed and describe the extended settings supported by the
same formulation.

\subsubsection{\rev{Feedback-Free Autoregressive Sequential Planning} under Opinion Dynamics}
\label{subsec:one_shot_planning}

\rev{Feedback-free autoregressive sequential planning} is an information constraint for opinion-dynamics-based
moderation. The planner may observe the initial graph--opinion instance,
including the pre-intervention settled expressed-opinion vector, but it may not
use newly recomputed settled expressed opinions after partial interventions as
inputs, rankings, or candidate-evaluation signals.

Let $\mathcal{I}_0=(G,\mathbf{s}^{(0)},\mathbf{z}^{(0)},k,\mathbf{c})$ denote
the initial instance, where $\mathbf{z}^{(0)}$ is the settled expressed-opinion
vector before any intervention, $k$ is the budget, and $\mathbf{c}$ is included
only in cost-aware settings.

\begin{definition}[\rev{Feedback-free autoregressive sequential planning} under opinion dynamics]
\label{def:one_shot_planning}
A planner satisfies the \rev{feedback-free planning regime} if it \rev{autoregressively constructs}
$\sigma=(v_1,\ldots,v_k)$ from $\mathcal{I}_0$ and deterministic bookkeeping
variables induced by previous selections, without using recomputed
intermediate settled-opinion feedback.

At step $t$, the planner may use the prefix $\sigma_{t-1}$, selected set
$T_{t-1}$, intervention marks, feasibility masks, costs, deterministic
mechanism-specific updates, and, for topology-modifying extensions, the
deterministic residual graph. It may also use the initial settled expressed
opinions $\mathbf{z}^{(0)}$ because they are part of $\mathcal{I}_0$. However,
it may not use intermediate settled-opinion vectors
$\mathbf{z}^{(1)}_{\mathcal{M}},\ldots,\mathbf{z}^{(k-1)}_{\mathcal{M}}$,
intermediate residual-polarization scores
$RPS_{\mathcal{M}}(\sigma_1),\ldots,RPS_{\mathcal{M}}(\sigma_{k-1})$, or
candidate scores obtained by temporarily applying a candidate action and
re-solving the opinion dynamics.
\end{definition}

Thus, \rev{feedback-free autoregressive planning} is not the same as a fixed static ranking. A method may
adapt to its own previous selections through deterministic bookkeeping, such as
marks, masks, remaining feasible nodes, costs, or residual graph updates. What
is excluded is feedback from re-equilibrating the opinion system after partial
interventions. For example, a heuristic that ranks nodes once using the initial
$\mathbf{z}^{(0)}$ satisfies the \rev{feedback-free regime}, whereas a select--recompute--
select method that repeatedly updates its ranking from
$\mathbf{z}^{(t)}_{\mathcal{M}}$ does not. Similarly, a learning-based greedy
method that requires recomputed expressed opinions after each selected prefix
as policy input or candidate-evaluation input is treated as full-information.

This restriction is both computational and methodological. Even in linear FJ
dynamics, repeated re-equilibration can be expensive on large graphs. In richer
opinion models, the dynamics may be nonlinear, lack a closed-form equilibrium,
or admit multiple limiting clusters or configurations, as in bounded-confidence,
biased assimilation, or antagonistic-interaction models
\cite{hegselmann2002opinion,deffuant2000mixing,dandekar2013biased,altafini2013consensus}.
Repeated intervene--re-equilibrate--replan procedures may therefore be costly
and may not yield a unique comparable intermediate state.

\begin{problem}[\rev{Feedback-free} ANP polarization moderation]
\label{prob:one_shot_anp}
Given an initial instance $\mathcal{I}_0$, an intervention mechanism
$\mathcal{M}\in\{\textsc{MI},\textsc{ME}\}$, and budget $k$, find an ordered
sequence $\sigma=(v_1,\ldots,v_k)\in\mathcal{P}_k(V)$ that minimizes
$ANP_{\mathcal{M}}(\sigma)$ subject to the \rev{feedback-free planning constraint} in
Definition~\ref{def:one_shot_planning}. In cost-aware settings, the objective
is replaced by $ANP_{\mathcal{M}}^{cost}(\sigma)$.
\end{problem}

Definition~\ref{def:one_shot_planning} rules out full-information
select--recompute--select procedures and candidate-evaluation heuristics that
test feasible nodes by temporary intervention and re-solving. In the standard
experiments, PACIFIER and all baselines without full-information feedback
satisfy this regime. Methods with suffix \textbf{-FI}, and methods such as
GNN-GreedyExt when they consume recomputed intermediate expressed opinions, are
evaluated separately as full-information counterparts in
Sec.~\ref{subsec:synthetic_fullinfo} and Sec.~\ref{subsec:runtime_both} to
measure the quality and efficiency trade-off of removing intermediate
equilibrium feedback.

\subsubsection{Extended Settings}
\label{subsec:extensions}

Beyond canonical \textsc{MI} and \textsc{ME}, PACIFIER evaluates three
extensions under the same ANP objective and \rev{feedback-free planning regime}. In
cost-aware moderation, each node $v$ has a nonnegative cost $c(v)$, and the
sequence is evaluated by the cost-aware ANP in Eq.~\eqref{eq:cost_anp}. In
continuous-opinion moderation, internal opinions lie in $[-1,1]$ rather than
$\{-1,+1\}$, while the FJ equilibrium, polarization index, and MI/ME semantics
remain unchanged. In topology-modifying node removal, selecting node $v_t$
removes it and all incident edges, i.e., $G_t=G_{t-1}\setminus\{v_t\}$, and the
FJ equilibrium and polarization index are computed on the residual graph. Thus,
the extensions differ only in the action-induced transition: costs reweight
actions, continuous opinions enrich the opinion space, and node removal changes
the graph topology.

\section{PACIFIER Framework}
\label{sec:pacifier}

\begin{figure*}[t]
    \centering
    \includegraphics[width=0.90\textwidth]{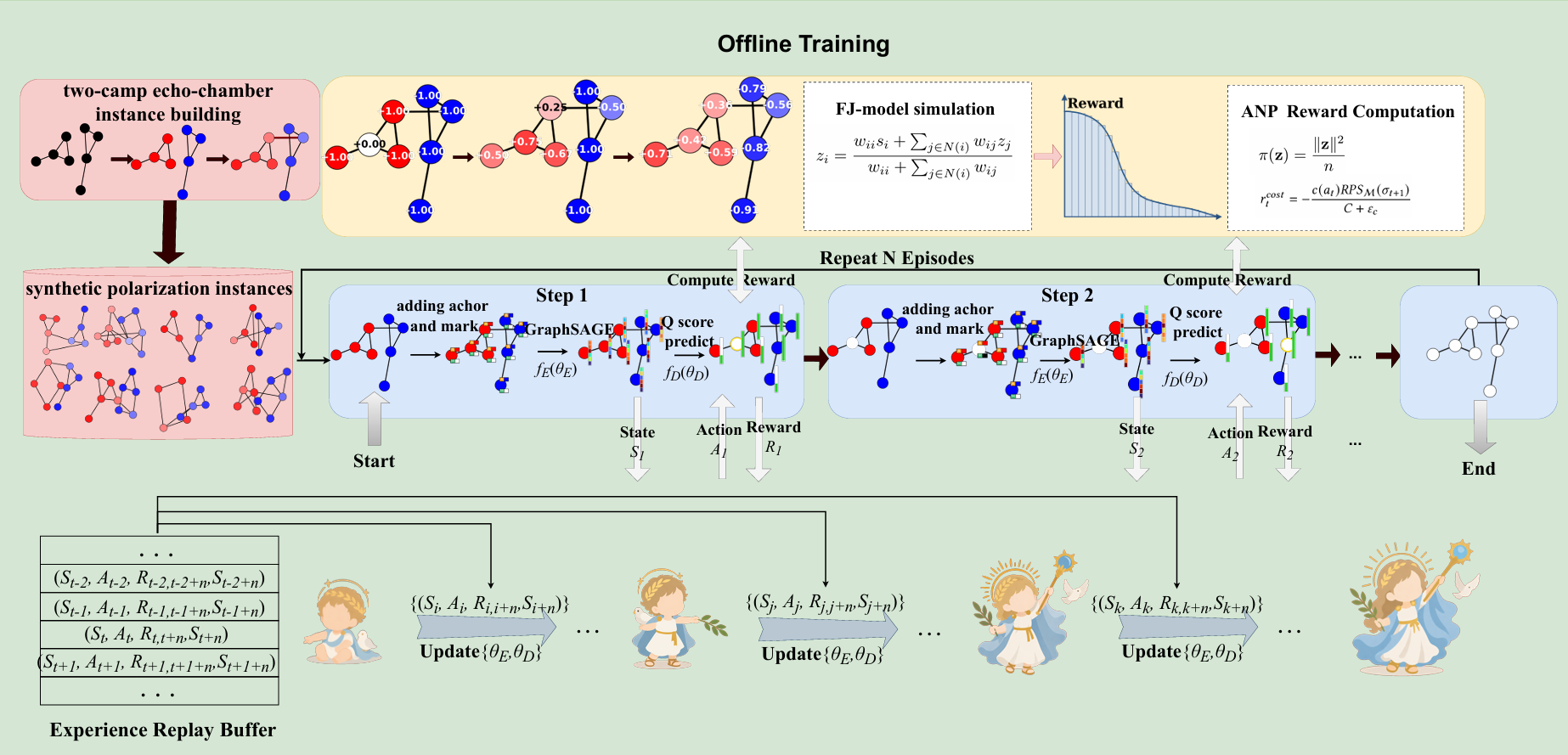}
    \par\nointerlineskip
    \hspace*{0.26em}%
    \includegraphics[width=0.90\textwidth]{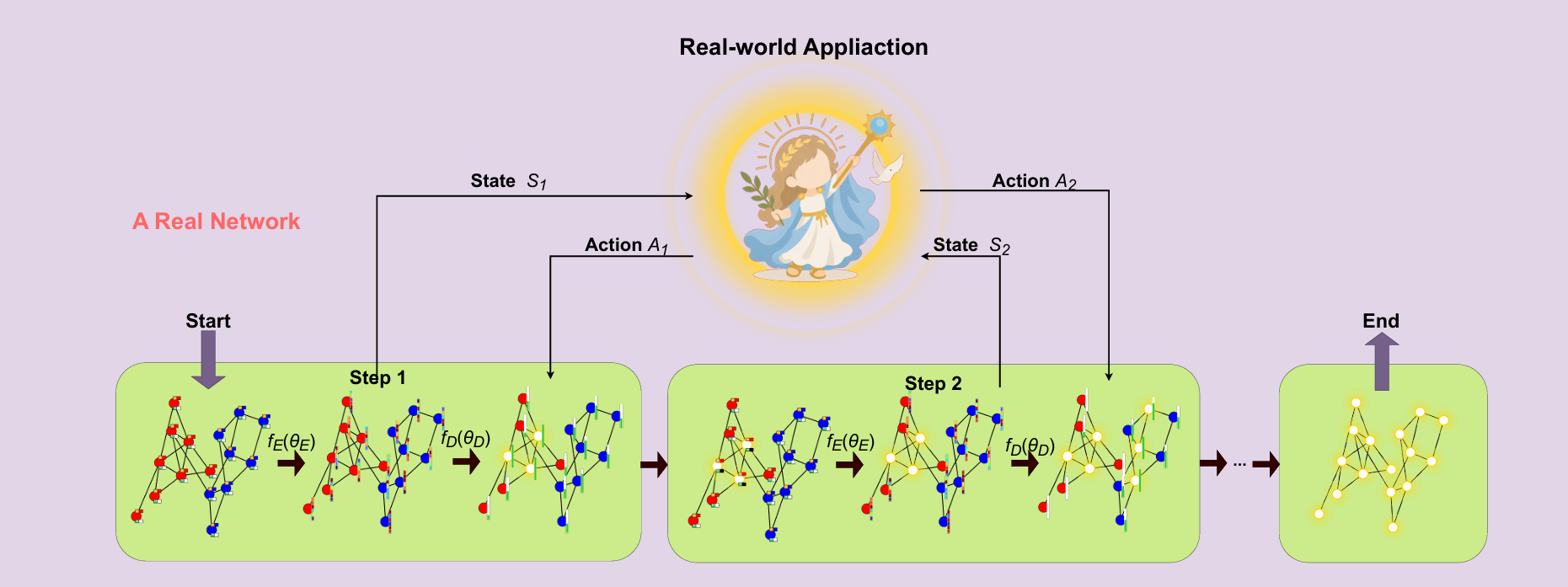}
    \vspace{-0.5em}
    \caption{PACIFIER workflow: offline training and real-world application.}
    \label{fig:pacifier_workflow}
    \label{fig:offline_training}
    \label{fig:online_application}
    \vspace{-0.9em}
\end{figure*}

This section presents \textbf{\textsc{PACIFIER}: 
\underline{P}\underline{A}\underline{C}ing Op\underline{I}nion Depolarization via a 
Uni\underline{F}\underline{I}\underline{E}d Graph Learning F\underline{R}amework}. As shown in
Fig.~\ref{fig:pacifier_workflow}, PACIFIER separates offline training from
\rev{feedback-free application}. During training, the environment simulates FJ-based
polarization after selected interventions and computes ANP-aligned rewards.
During application, the learned policy scores feasible nodes by feed-forward
inference and updates only deterministic bookkeeping variables, without using
recomputed post-intervention expressed opinions as intermediate policy input.

Following the modular design of graph-learning intervention frameworks,
PACIFIER consists of four components:
\begin{enumerate}[leftmargin=*, topsep=2pt, itemsep=1pt]
    \item \textbf{Two-Camp Echo-Chamber Building}, which generates small
    polarized graph--opinion instances for cross-scale training;
    \item \textbf{History-Aware Encoder--Decoder with Polarization Features},
    which maps the current moderation state to node scores using intervention
    history, graph embeddings, and scale-normalized polarization signals;
    \item \textbf{FJ Polarization Simulation and ANP Reward Computation}, which
    applies the selected intervention, recomputes the FJ equilibrium for
    reward/evaluation, and produces residual-polarization rewards;
    \item \textbf{RL and Greedy Training Process}, which trains
    PACIFIER-RL for long-horizon action values and PACIFIER-Greedy for
    immediate action ranking under the same node-action interface.
\end{enumerate}

\subsection{Two-Camp Echo-Chamber Building}
\label{sec:pacifier_data}

We train PACIFIER on a synthetic two-camp echo-chamber distribution
$(G,\mathbf{s}^{(0)},\mathbf{c})\sim\mathcal{D}_{\mathrm{echo}}$,
which is designed to expose the policy to the mechanism that drives FJ
polarization: degree-heterogeneous camps, sparse cross-camp exposure, and
camp-aligned opinions. \rev{The goal is not to imitate a target network, but to learn a scale-stable
policy that transfers directly, without target-specific retraining, from small
synthetic graphs to unseen real networks with a comparable two-echo-chamber
polarization structure.}

Each training instance is built from two approximately balanced
Barab\'asi--Albert (BA) graphs~\cite{barabasi1999emergence}. For canonical
\textsc{MI}/\textsc{ME}, the graph size is sampled from $[18,50]$; for node
removal, it is sampled from $[30,50]$. The two BA graphs form camps $V^+$ and
$V^-$, their within-camp edges are preserved, and a small number of random
cross-camp edges is added to create sparse exposure between opposing camps.
For binary-opinion settings, $s_i^{(0)}=+1$ for $i\in V^+$ and
$s_i^{(0)}=-1$ for $i\in V^-$. Continuous-opinion settings preserve the
camp-aligned sign while sampling bounded magnitudes in $[-1,1]$, and
cost-aware settings additionally sample node costs $\mathbf{c}$. In unweighted
or non-cost settings, costs are set to a constant.

\subsection{History-Aware Encoder--Decoder with Polarization Features}
\label{sec:pacifier_encoding_policy}

At step $t$, PACIFIER represents the moderation state as
\begin{equation}
\label{eq:pacifier_mdp_state}
\mathcal{S}_t=(G_t,X_t,\mathbf{u}_t,\mathbf{m}_t),
\end{equation}
where $G_t=G$ for fixed-topology \textsc{MI}/\textsc{ME}, and $G_t$ is the
residual graph for node removal. The feasible action set is
\begin{equation}
\label{eq:pacifier_mdp_action}
\mathcal{A}_t=\{v\in V_t:m_t(v)=1\}.
\end{equation}

The key observability issue is that \textsc{MI} and \textsc{ME} preserve the
graph topology, so adjacency alone does not reveal which nodes have already
been intervened on. PACIFIER therefore uses history-aware node features:
\begin{equation}
\label{eq:pacifier_node_feat}
\mathbf{x}_t(v)
=
[
\phi_t(v),\;
\phi_0(v),\;
\mathrm{mark}_t(v),\;
c(v)
],
\qquad
\mathrm{mark}_t(v)=\mathbb{I}[v\in T_t].
\end{equation}
Here $\phi_t(v)$ is the current mechanism-specific opinion attribute,
$\phi_0(v)=s^{(0)}(v)$ is the initial-opinion anchor,
$\mathrm{mark}_t(v)$ records whether node $v$ has been selected, and $c(v)$ is
an optional cost. For \textsc{MI}, $\phi_t(v)=s^{(t)}(v)$ and selected nodes
have $s^{(t)}(v)=0$. For \textsc{ME}, selected nodes are represented by the
intervention mark and treated as neutral expressed-opinion boundary nodes in
the simulation module. The feasibility mask is
$m_t(v)=1-\mathrm{mark}_t(v)$.

The anchor and mark make the fixed-topology state history-aware:
$\phi_0$ distinguishes natural neutrality from intervention-induced neutrality,
while $\mathrm{mark}_t$ explicitly records the selected set. Together, they
resolve the state-aliasing problem caused by topology-preserving
continuous-state interventions.

The node-feature matrix $X_t$ is first projected into initial node embeddings:
\begin{equation}
\label{eq:pacifier_initial_embedding}
\mathbf{h}_v^{0}
=
\operatorname{Normalize}
\left(
\operatorname{ReLU}(W_0\mathbf{x}_t(v))
\right),
\qquad v\in V_t.
\end{equation}

We also compute a fixed-dimensional polarization feature vector
\begin{equation}
\label{eq:pacifier_aux_feat}
\mathbf{u}_t=
\left[
\frac{|T_t|}{n},\;
\frac{|E_{\mathrm{cov}}(t)|}{|E_t|+\epsilon_m},\;
\frac{|E_{\pm}(t)|}{|E_t|+\epsilon_m},\;
\frac{T(t)}{n^2+\epsilon_n},\;
\frac{T^+(t)}{n^2+\epsilon_n},\;
\frac{T^-(t)}{n^2+\epsilon_n}
\right].
\end{equation}
Here $E_{\mathrm{cov}}(t)$ is the set of edges incident to selected nodes,
$E_{\pm}(t)$ is the set of uncovered edges connecting active nodes with
opposite opinion signs, $T(t)$ is the active two-hop count, and
$T^+(t),T^-(t)$ are the corresponding two-hop counts inside the positive and
negative active opinion groups. These features summarize intervention progress,
covered edges, remaining cross-camp exposure, and residual within-camp cohesion
using scale-normalized ratios.

The encoder is a GraphSAGE-style message-passing network
\cite{hamilton2017inductive} with a virtual super-node $v_\star$. Real nodes
aggregate from graph neighbors, while $v_\star$ aggregates from all real nodes
and provides a graph-level representation. After $L_{\mathrm{mp}}$ layers, the
encoder returns node embeddings $\mathbf{h}_v$ and graph embedding
$\mathbf{g}_t=\mathbf{h}_{v_\star}^{L_{\mathrm{mp}}}$.

Given $\mathbf{h}_v$, $\mathbf{g}_t$, and $\mathbf{u}_t$, the decoder predicts
a score for each node:
\begin{equation}
\label{eq:pacifier_q_value}
q_\theta(\mathcal{S}_t,v)
=
f_D
\left(
[
W_b\operatorname{vec}(\mathbf{h}_v\mathbf{g}_t^\top)
\Vert
\mathbf{u}_t
];
\theta_D
\right),
\qquad
\theta=\{\theta_E,\theta_D\}.
\end{equation}
Previously selected nodes are excluded by masked scoring:
\begin{equation}
\label{eq:pacifier_masked_q}
\widetilde{q}_\theta(\mathcal{S}_t,v)
=
q_\theta(\mathcal{S}_t,v)+(1-m_t(v))(-C_{\mathrm{mask}}),
\end{equation}
where $C_{\mathrm{mask}}$ is a sufficiently large constant. Thus, this module
takes the current graph--opinion state as input and outputs a masked action
score for every candidate node.

\subsection{FJ Polarization Simulation and ANP Reward Computation}
\label{sec:pacifier_simulation}

The simulation module defines what happens after a node is selected and how the
resulting prefix is evaluated. Given a selected action $a_t$, PACIFIER updates
the deterministic bookkeeping variables and then applies the mechanism-specific
transition. For \textsc{MI}, the selected node's internal opinion is set to
zero. For \textsc{ME}, the selected node is fixed to neutral expressed opinion
as a boundary condition. For node removal, the selected node and all incident
edges are removed from the residual graph.

After applying the prefix $\sigma_{t+1}$, the environment computes the settled
expressed opinion
\[
\mathbf{z}^{(t+1)}_{\mathcal{M}}
=
\operatorname{Eq}_{\mathcal{M}}(G_{t+1},\mathbf{s}^{(0)},\sigma_{t+1}),
\]
and evaluates residual polarization by
\begin{equation}
\label{eq:pacifier_sim_rps}
RPS_{\mathcal{M}}(\sigma_{t+1})
=
\pi_{G_{t+1}}\!\left(\mathbf{z}^{(t+1)}_{\mathcal{M}}\right).
\end{equation}
The reward is the negative residual polarization normalized by the cardinality
progress:
\begin{equation}
\label{eq:pacifier_reward_unweighted}
r_t
=
-
\frac{
RPS_{\mathcal{M}}(\sigma_{t+1})
}{
|V|
}.
\end{equation}
For cost-aware moderation, with $C=\sum_{v\in V}c(v)$, the reward becomes
\begin{equation}
\label{eq:pacifier_reward_cost}
r_t^{cost}
=
-
\frac{
c(a_t)RPS_{\mathcal{M}}(\sigma_{t+1})
}{
C+\varepsilon_c
}.
\end{equation}
The factor $RPS_{\mathcal{M}}(\emptyset)+\varepsilon$ is used only for
ANP-based evaluation across datasets with different initial polarization
levels, while training rewards instead use node- or cost-progress
normalization to preserve absolute polarization differences.

This simulation module is used during offline training to produce rewards and
during evaluation to compute ANP. It is not used as intermediate policy input
during \rev{feedback-free application}. Therefore, PACIFIER can learn from FJ-equilibrium
feedback in simulation while still deploying without select--recompute--select
replanning.

\subsection{RL and Greedy Training Process}
\label{sec:pacifier_training}

PACIFIER uses the same encoder--decoder scoring interface for both its RL and
Greedy variants. PACIFIER-RL learns long-horizon action values with $n$-step
Q-learning. For $R_{t:t+n}=\sum_{i=0}^{n-1}\gamma^i r_{t+i}$, the target is
\begin{equation}
\label{eq:pacifier_td_target}
y_t
=
R_{t:t+n}
+
\gamma^n
\max_{v:m_{t+n}(v)=1}
q_{\theta^-}(\mathcal{S}_{t+n},v),
\end{equation}
and the loss is
\begin{equation}
\label{eq:pacifier_rl_loss}
\mathcal{L}_{\mathrm{RL}}(\theta)
=
\mathbb{E}
\left[
\left(
q_\theta(\mathcal{S}_t,a_t)-y_t
\right)^2
\right].
\end{equation}
PACIFIER-Greedy is the myopic, non-bootstrapped counterpart of PACIFIER-RL and learns immediate action scores:
\begin{equation}
\label{eq:pacifier_greedy_loss}
\mathcal{L}_{\mathrm{Greedy}}(\theta)
=
\mathbb{E}
\left[
\left(
q_\theta(\mathcal{S}_t,a_t)-r_t
\right)^2
\right].
\end{equation}

Algorithm~\ref{alg:pacifier} summarizes the complete process. During offline
training, PACIFIER samples synthetic polarized instances, scores feasible nodes,
applies selected interventions, computes ANP-aligned rewards through the
simulation module, and updates the encoder--decoder parameters. During \rev{feedback-free}
application, the trained model constructs an ordered intervention sequence by
feed-forward scoring and deterministic bookkeeping only.

\begin{algorithm}[t]
\caption{PACIFIER Training and \rev{Feedback-Free Application}}
\label{alg:pacifier}
\footnotesize
\setlength{\algorithmicindent}{0.8em}
\begin{algorithmic}[1]
\REQUIRE Mechanism $\mathcal{M}$, horizon $k$, mode
$\eta\in\{\mathrm{RL},\mathrm{Greedy}\}$, training distribution
$\mathcal{D}_{\mathrm{echo}}$
\ENSURE Learned parameters $\theta$ and \rev{feedback-free} sequence $\sigma$

\STATE \textbf{Offline training:} initialize encoder--decoder parameters
$\theta=\{\theta_E,\theta_D\}$ and target parameters $\theta^-$ for RL.
\FOR{each training episode}
    \STATE Sample $(G,\mathbf{s}^{(0)},\mathbf{c})\sim\mathcal{D}_{\mathrm{echo}}$.
    \STATE Initialize $T_0=\emptyset$, $\sigma_0=\emptyset$, and
    $\mathcal{S}_0=(G_0,X_0,\mathbf{u}_0,\mathbf{m}_0)$.
    \FOR{$t=0,\ldots,k-1$}
        \STATE Encode $\mathcal{S}_t$ and score feasible nodes by
        $\widetilde{q}_\theta(\mathcal{S}_t,v)$.
        \STATE Select training action $a_t$.
        \STATE Apply $a_t$ under $\mathcal{M}$ and update deterministic
        bookkeeping.
        \STATE Compute $RPS_{\mathcal{M}}(\sigma_{t+1})$ and reward $r_t$.
        \STATE Build
        $\mathcal{S}_{t+1}=(G_{t+1},X_{t+1},\mathbf{u}_{t+1},\mathbf{m}_{t+1})$.
        \IF{$\eta=\mathrm{RL}$}
            \STATE Update $\theta$ using the $n$-step target
            $y_t$ and $\mathcal{L}_{\mathrm{RL}}$.
        \ELSE
            \STATE Update $\theta$ using immediate reward target
            $r_t$ and $\mathcal{L}_{\mathrm{Greedy}}$.
        \ENDIF
    \ENDFOR
\ENDFOR

\STATE \textbf{\rev{Feedback-free application:}} initialize $\sigma=\emptyset$ and
$\mathcal{S}_0$ from the target instance.
\FOR{$t=0,\ldots,k-1$}
    \STATE Select
    $a_t=\arg\max_{v:m_t(v)=1}q_\theta(\mathcal{S}_t,v)$.
    \STATE Append $a_t$ to $\sigma$.
    \STATE Update marks, masks, auxiliary features, and mechanism-specific
    attributes.
    \STATE For node removal, set $G_{t+1}=G_t\setminus\{a_t\}$; otherwise keep
    $G_{t+1}=G_t$.
\ENDFOR
\RETURN $\sigma$
\end{algorithmic}
\end{algorithm}

The mechanism only changes the transition and reward evaluation. Cost-aware
moderation reweights actions, continuous-opinion moderation changes the opinion
space, and node removal changes graph topology. Across all settings, PACIFIER
keeps the same pipeline: two-camp instance building, history-aware
encoder--decoder scoring, FJ-based polarization simulation, and RL/Greedy
training.

\rev{PACIFIER-RL uses 5-step Q-learning ($\gamma=1$), $\epsilon$-greedy
exploration with $\epsilon$ linearly decayed from $1.0$ to $0.05$ over 10,000
iterations (then fixed at $0.05$), Adam with learning rate $10^{-4}$, batch size
64, replay size $5\times10^5$, and target-network updates every 1,000 iterations.}
Implementation details,
including the exploration strategy and training hyperparameters, are available in
the released source code. Training-convergence curves for both PACIFIER variants
are also provided in the released source code.

\section{Experiments}
\label{sec:experiments}
\setlist[itemize]{leftmargin=*, topsep=0pt, itemsep=0pt, parsep=0pt, partopsep=0pt}

\begin{figure*}[t]
\centering
\includegraphics[width=0.98\textwidth]{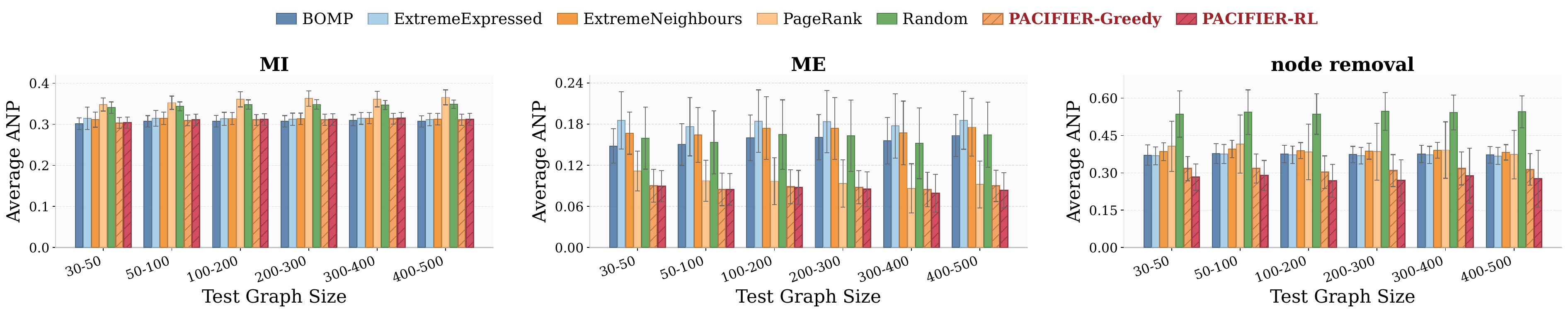}
\vspace{-1.1em}
\caption{Synthetic cross-scale results on \textsc{MI}, \textsc{ME}, and node removal; bars show ANP, lower is better.}
\label{fig:synth_bars_all}
\vspace{-1em}
\end{figure*}


\begin{figure}[!tbp]
\centering
\includegraphics[width=0.98\linewidth]{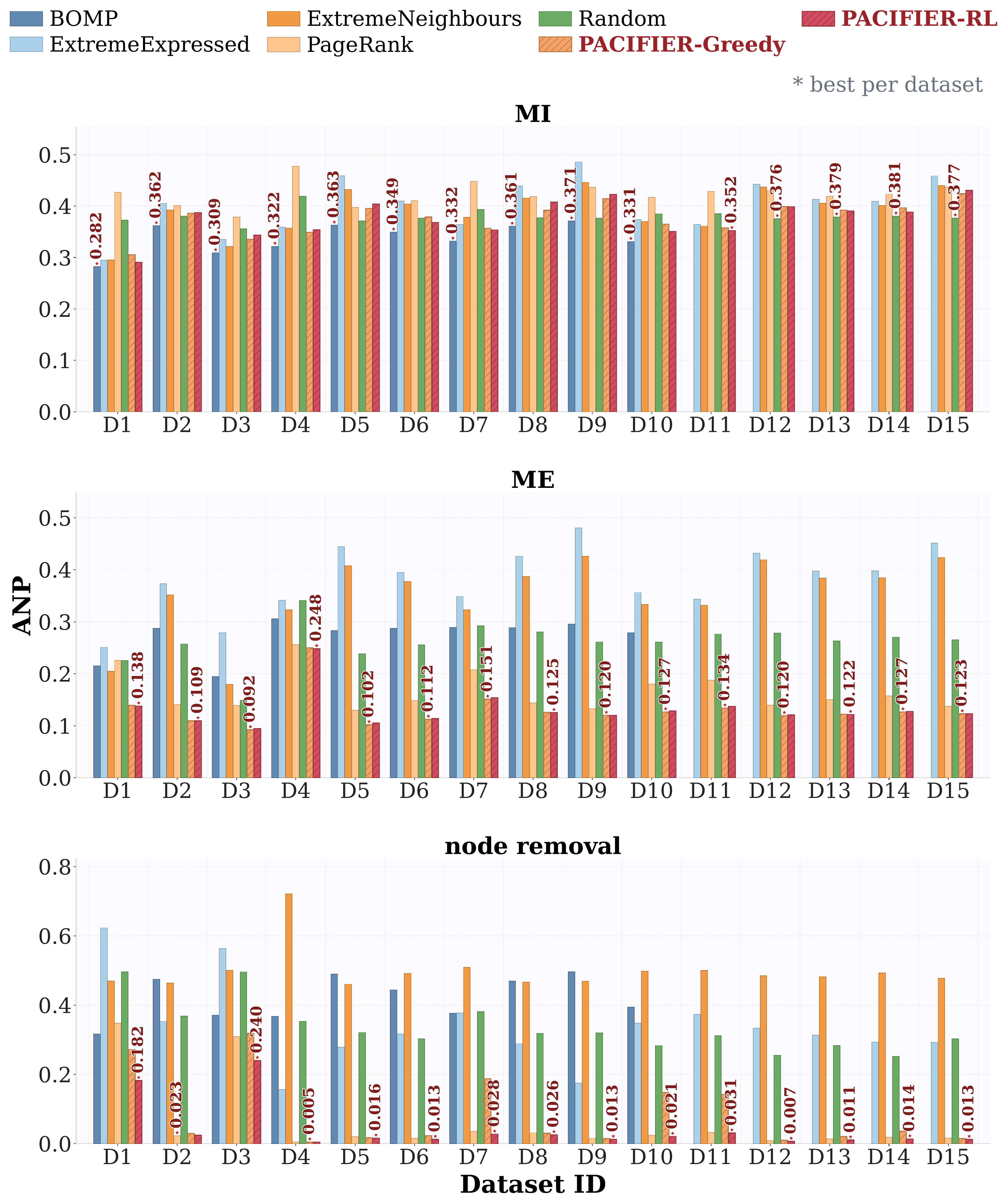}
\vspace{-0.5em}
\caption{
Real-world per-dataset ANP comparison on \textsc{MI}, \textsc{ME}, and node removal.
}
\label{fig:real_canonical_bars_all}
\label{fig:real_mi_bar}
\label{fig:real_me_bar}
\label{fig:real_node_removal_bar}
\vspace{-1em}
\end{figure}

\begin{figure}[!tbp]
\centering
\includegraphics[width=0.98\linewidth]{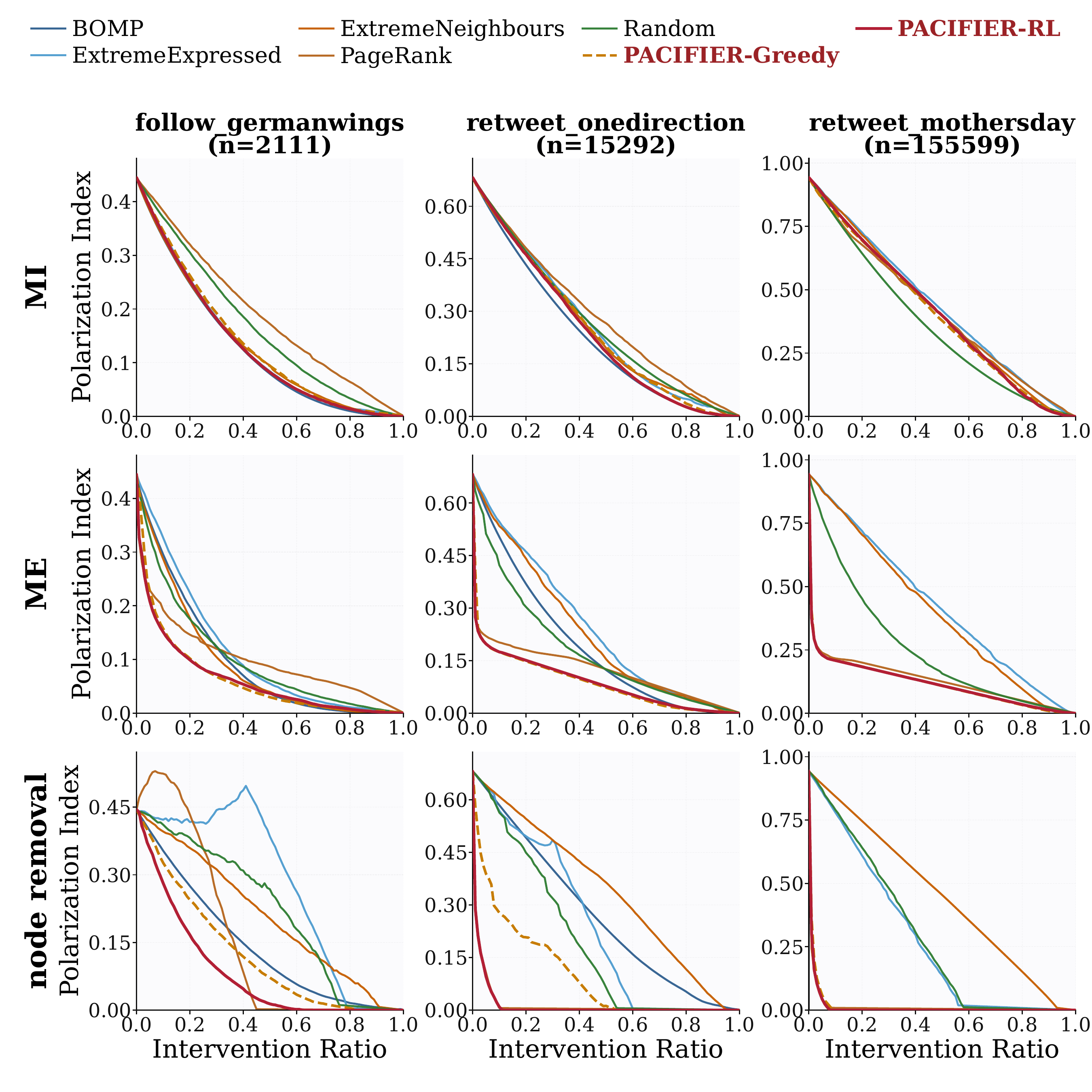}
\vspace{-0.9em}
\caption{Real-world representative polarization trajectories on \textsc{MI}, \textsc{ME}, and node removal.}
\label{fig:real_traj_3x3_all}
\label{fig:real_mi_traj_rep}
\label{fig:real_me_traj_rep}
\label{fig:real_node_removal_traj_rep}
\vspace{-2.1em}
\end{figure}

\begin{figure}[!tbp]
\centering

\begin{subfigure}[t]{\linewidth}
    \centering
    \includegraphics[width=0.98\linewidth]{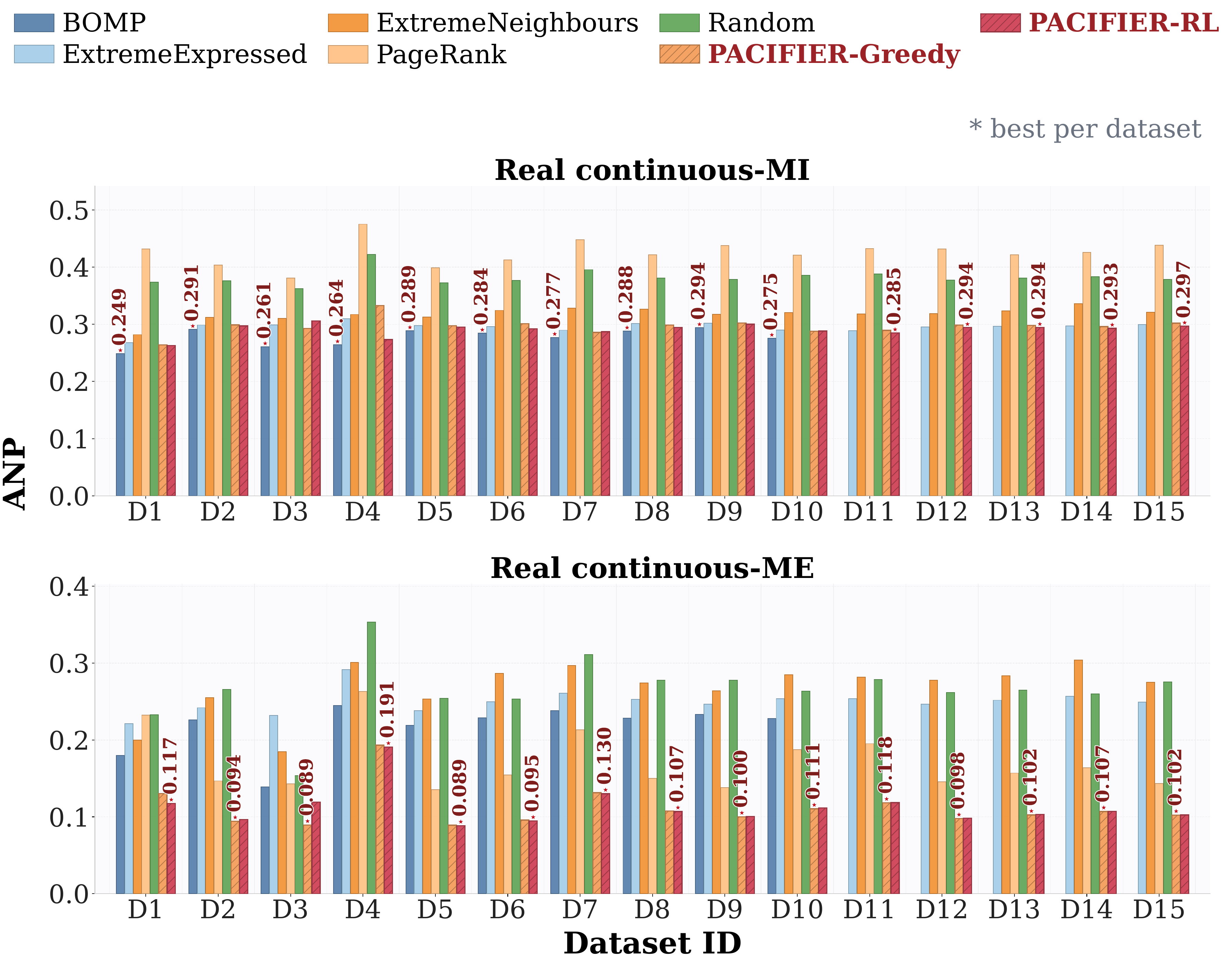}
    \caption{Continuous-opinion setting.}
    \label{fig:real_continuous_bars_all}
    \label{fig:real_continuous_mi_bar}
    \label{fig:real_continuous_me_bar}
\end{subfigure}

\vspace{0.1em}

\begin{subfigure}[t]{\linewidth}
    \centering
    \includegraphics[width=0.98\linewidth]{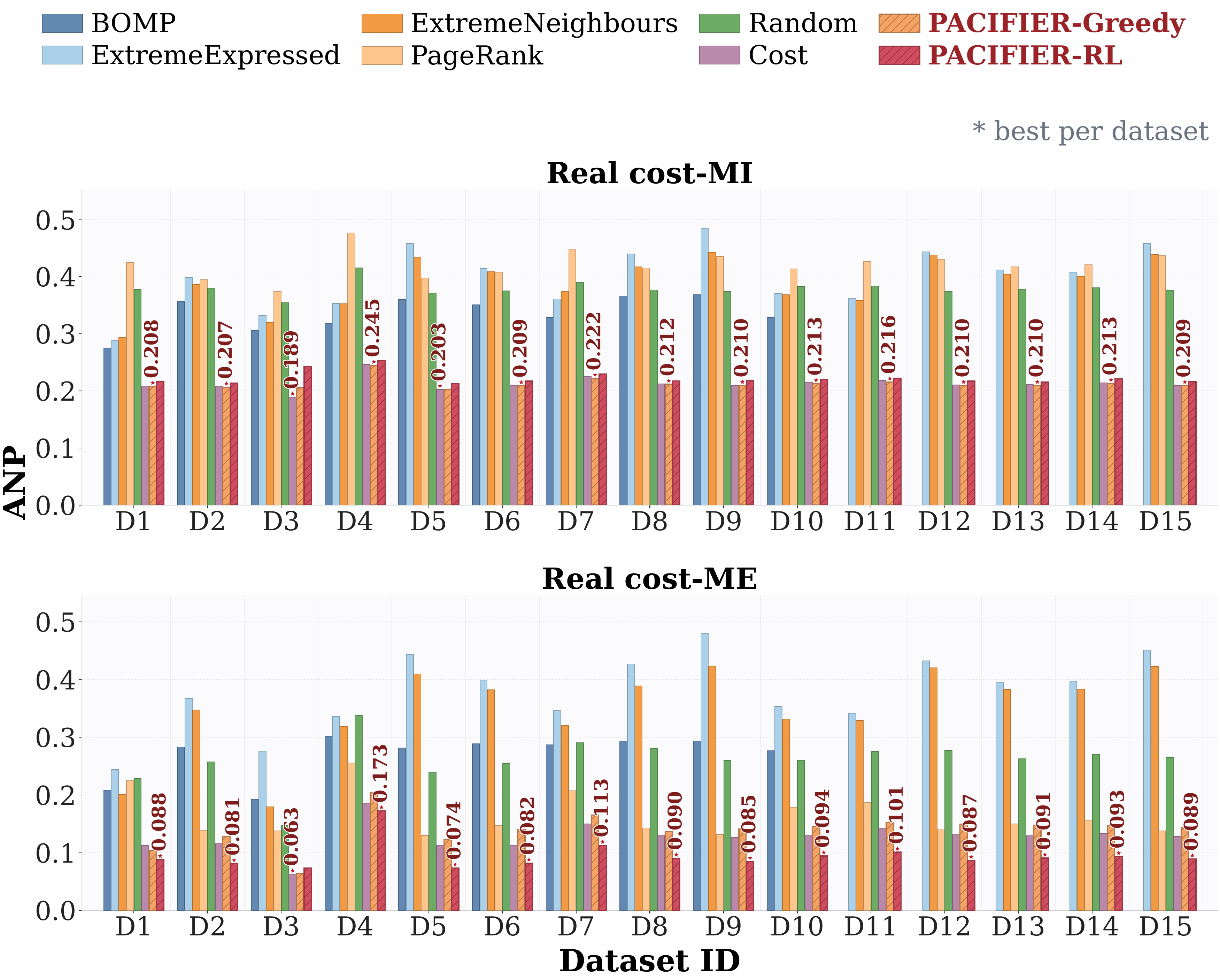}
    \caption{Cost-aware setting.}
    \label{fig:real_cost_bars_all}
    \label{fig:real_cost_mi_bar}
    \label{fig:real_cost_me_bar}
\end{subfigure}

\vspace{-0.4em}
\caption{
Extended real-world ANP comparison on continuous-opinion and cost-aware settings.
}
\label{fig:real_extended_bars_all}
\vspace{-1em}
\end{figure}

We evaluate PACIFIER on synthetic and real-world polarized networks. Unless
otherwise specified, all methods are evaluated by accumulated normalized
polarization (ANP), where lower values indicate earlier and more persistent
polarization reduction. The experiments examine cross-scale transfer,
comparison with baselines, the value of RL, the cost and benefit
of full-information feedback, runtime scalability, and representation ablations.
For the real-data experiments, all methods evaluate polarization after every
1\% node increment before proceeding.

\subsection{Experimental Setup}
\label{subsec:experimental_setup}

\paragraph{Datasets.}
The real-world datasets are derived from the Twitter topic graph collection of
Garimella et al.~\cite{garimella2018quantifying}. We remove graphs with
self-loops or disconnected structure, obtaining 31 structurally valid graphs.
Using the original two-camp partition, we assign $s_i^{(0)}=+1$ to one camp and
$s_i^{(0)}=-1$ to the other, and retain graphs with
$\pi(\mathbf{z}^{(0)})>0.4$. This yields 15 polarized real-world benchmark
graphs, summarized in Table~\ref{tab:filtered_datasets_pol04}. Dataset IDs
D1--D15 follow ascending node count and are used as the per-dataset x-axis
labels in our real-world bar figures.

\begin{table}[t]
\centering
\caption{
Datasets retained after filtering by $\pi(\mathbf{z}^{(0)})>0.4$.
}
\label{tab:filtered_datasets_pol04}
\scriptsize
\setlength{\tabcolsep}{1.7pt}
\renewcommand{\arraystretch}{1.00}
\resizebox{\linewidth}{!}{
\begin{tabular}{clrr|clrr}
\toprule
ID & Dataset & $n$ & $\pi(\mathbf{z}^{(0)})$ &
ID & Dataset & $n$ & $\pi(\mathbf{z}^{(0)})$ \\
\midrule
D1  & follow\_germanwings         & 2111   & 0.445 &
D9  & retweet\_wcw                & 10674  & 0.990 \\
D2  & retweet\_russia\_march      & 2134   & 0.922 &
D10 & retweet\_onedirection       & 15292  & 0.682 \\
D3  & follow\_mothersday          & 2225   & 0.606 &
D11 & retweet\_leadersdebate      & 25983  & 0.663 \\
D4  & retweet\_nationalkissingday & 4638   & 0.496 &
D12 & retweet\_jurassicworld      & 26407  & 0.911 \\
D5  & retweet\_ff                 & 5401   & 0.984 &
D13 & retweet\_germanwings        & 29763  & 0.843 \\
D6  & retweet\_gunsense           & 7106   & 0.835 &
D14 & retweet\_nepal              & 40579  & 0.807 \\
D7  & retweet\_ultralive          & 9261   & 0.596 &
D15 & retweet\_mothersday         & 155599 & 0.943 \\
D8  & retweet\_sxsw               & 9304   & 0.930 &
    &                               &        &       \\
\bottomrule
\end{tabular}
}
\vspace{-0.6em}
\end{table}

\paragraph{Implementation.}
All experiments were conducted on a single desktop workstation with an Intel
Core i5-12400F CPU, an NVIDIA GeForce RTX 3060 Ti GPU, and 32 GB RAM. The
implementation was developed in Python. PACIFIER uses GPU acceleration for
neural inference on large real-world graphs, while analytical baselines and
FJ-equilibrium evaluation run on CPU.

\paragraph{Baselines.}
We compare PACIFIER with analytical, structural, opinion-aware, cost-aware,
learning-based, and oracle-style baselines. \textbf{PACIFIER-RL} learns
long-horizon action values, while \textbf{PACIFIER-Greedy} learns immediate
action scores. \textbf{BOMP} is the
main analytical baseline for linear \textsc{MI}~\cite{matakos2017measuring}.
Due to its high computational cost, BOMP is evaluated only on datasets up to D10
(\texttt{retweet\_onedirection}, $n=15{,}292$). Other \rev{feedback-free} baselines include
\textbf{ExtremeExpressed} and \textbf{ExtremeNeighbours}, two heuristics from
Matakos et al.~\cite{matakos2017measuring} that rank nodes by expressed-opinion
extremeness and neighborhood extremeness, respectively; \textbf{PageRank}~\cite{page1999pagerank};
\textbf{Random}, which selects nodes uniformly at random; and, in cost-aware
tasks, \textbf{Cost}, which prioritizes low-cost nodes. Methods with suffix
\textbf{-FI} may recompute intermediate settled-opinion states during selection.
The oracle \textbf{Greedy} baseline exhaustively tries every feasible candidate
at each step. We also include \textbf{GNN-GreedyExt}, an aligned reproduction of
the GNN-assisted greedy method of Mylonas and Spyropoulos~\cite{mylonas2026opinion},
trained and evaluated under the same synthetic protocol as PACIFIER.

\subsection{Cross-scale Generalization}
\label{subsec:exp_cross_scale}

We first evaluate cross-scale transfer on synthetic two-camp echo-chamber graphs.
PACIFIER is trained on small graphs and tested on six size ranges:
$30$--$50$, $50$--$100$, $100$--$200$, $200$--$300$, $300$--$400$, and
$400$--$500$, with 100 graphs per range.

Figure~\ref{fig:synth_bars_all} supports three findings. Across all tested
size ranges, from $30$--$50$ to $400$--$500$ nodes, PACIFIER maintains stable
performance, indicating robust small-to-large transfer.
\begin{itemize}
\setlength{\itemsep}{0pt}
\setlength{\parskip}{0pt}
\setlength{\parsep}{0pt}
\setlength{\topsep}{1pt}
    \item \textsc{MI}: PACIFIER is only \textbf{1.01\%} worse than BOMP,
    while still outperforming the second-best non-PACIFIER baseline by
    \textbf{1.16\%}.
    \item \textsc{ME}: PACIFIER improves over PageRank, the strongest
    non-PACIFIER baseline, by \textbf{10.92\%}.
    \item \textit{node removal}: PACIFIER-RL improves over PageRank by
    \textbf{27.09\%}, showing the largest gain in the long-horizon structural
    setting.
\end{itemize}

\vspace{-0.25em}
\subsection{PACIFIER vs. Non-PACIFIER Baselines}
\label{subsec:exp_pacifier_vs_baselines}

We next compare PACIFIER with non-PACIFIER baselines on real-world datasets.
We consider canonical \textsc{MI}, canonical \textsc{ME}, and
\textit{node removal}, followed by continuous and cost-aware extensions.

\paragraph{Canonical real-world settings.}

Figures~\ref{fig:real_canonical_bars_all} and~\ref{fig:real_traj_3x3_all}
summarize the canonical real-world results. Percentage gains are computed
per dataset and then averaged across datasets.

\begin{itemize}
\setlength{\itemsep}{0pt}
\setlength{\parskip}{0pt}
\setlength{\parsep}{0pt}
\setlength{\topsep}{1pt}
    \item \textsc{MI}: BOMP remains strongest on the 10 comparable datasets,
    where PACIFIER trails it by \textbf{7.4\%}. Excluding BOMP, PACIFIER gives
    the lowest average ANP among all scalable methods, although the gains are
    modest.

    \item \textsc{ME}: PACIFIER consistently outperforms all baselines,
    improving over PageRank, the strongest non-PACIFIER baseline, by
    \textbf{21.1\%} on average.

    \item \textit{node removal}: PACIFIER also performs best, improving over
    PageRank by \textbf{19.2\%} and over ExtremeExpressed by \textbf{90.4\%}
    on average.
\end{itemize}

Overall, \textsc{MI} leaves relatively limited optimization room, whereas
\textsc{ME} and node removal show clearer gains for PACIFIER.

\paragraph{Extended real-world settings.}

As shown in Figs.~\ref{fig:real_continuous_bars_all}
and~\ref{fig:real_cost_bars_all}, the extended settings reinforce the same
conclusion. Percentage gains are computed per dataset and then averaged across
datasets.
\begin{itemize}
\setlength{\itemsep}{0pt}
\setlength{\parskip}{0pt}
\setlength{\parsep}{0pt}
\setlength{\topsep}{1pt}
    \item \textit{continuous opinions}: PACIFIER is competitive on
    continuous-\textsc{MI}, but still trails BOMP by \textbf{4.0\%} on the
    10 comparable datasets. On continuous-\textsc{ME}, PACIFIER improves over
    PageRank, the strongest non-PACIFIER baseline, by \textbf{35.3\%}.

    \item \textit{cost-aware moderation}: In cost-\textsc{MI}, PACIFIER is
    essentially tied with the Cost baseline, with a \textbf{0.08\%} lower
    per-dataset relative score. In cost-\textsc{ME}, PACIFIER improves over
    Cost, the strongest non-PACIFIER baseline, by \textbf{25.9\%}.
\end{itemize}

\vspace{-0.25em}
\subsection{When PACIFIER-RL Outperforms PACIFIER-Greedy}
\label{subsec:exp_rl_vs_greedy}

PACIFIER-RL is most useful when the intervention has long-horizon effects that
cannot be captured well by immediate action scores.
\begin{itemize}
\setlength{\itemsep}{0pt}
\setlength{\parskip}{0pt}
\setlength{\parsep}{0pt}
\setlength{\topsep}{1pt}
    \item Synthetic \textsc{ME}: in Fig.~\ref{fig:synth_bars_all}, the RL
    gain grows with graph size, from \textbf{0.00\%} in the 30--50 bucket to
    \textbf{6.99\%} in the 400--500 bucket, with an average gain of
    \textbf{2.80\%}.
    
    \item Synthetic \textit{node removal}: in Fig.~\ref{fig:synth_bars_all},
    PACIFIER-RL improves over PACIFIER-Greedy by \textbf{10.56\%} on average,
    showing the benefit of long-horizon planning in topology-changing
    interventions.
    
    \item Real-world long-horizon settings: in
    Figs.~\ref{fig:real_canonical_bars_all} and~\ref{fig:real_cost_bars_all},
    PACIFIER-RL substantially outperforms PACIFIER-Greedy in settings where
    delayed effects matter, improving by \textbf{37.69\%} in
    \textit{node removal} and \textbf{30.92\%} in cost-\textsc{ME}.
\end{itemize}

\begin{figure*}[!t]
\centering
\includegraphics[width=0.98\textwidth]{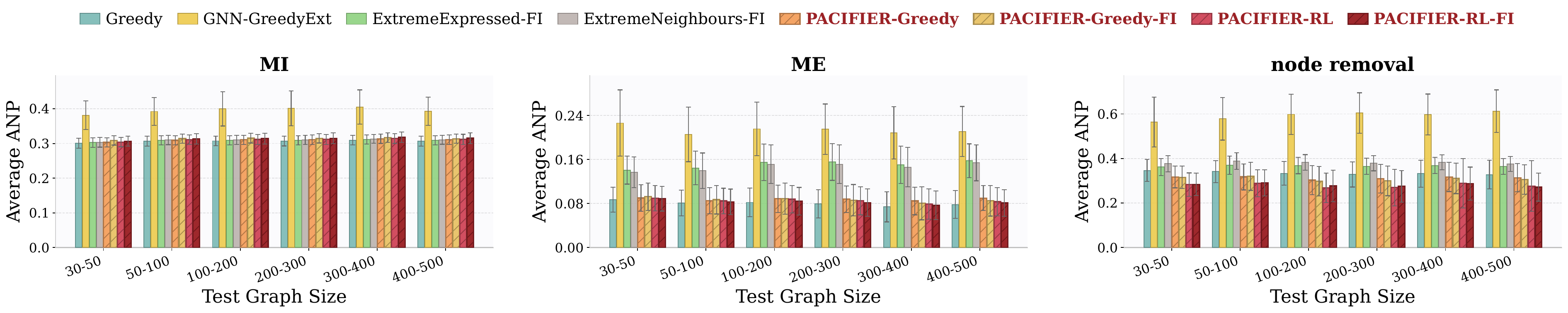}
\vspace{-0.5em}
\caption{Synthetic full-information comparison; FI methods may use intermediate equilibria.}
\label{fig:fullinfo_synth_all}
\vspace{-1em}
\end{figure*}

\subsection{Full-information Comparison}
\label{subsec:synthetic_fullinfo}

We compare standard \rev{feedback-free} PACIFIER with full-information variants,
oracle-style Greedy, and GNN-GreedyExt. FI methods may recompute intermediate
settled-opinion states during selection; standard PACIFIER cannot. All
learning-based methods use the same synthetic training and evaluation protocol.

Figure~\ref{fig:fullinfo_synth_all} shows that PACIFIER remains competitive even
against FI methods.
\begin{itemize}
\setlength{\itemsep}{0pt}
\setlength{\parskip}{0pt}
\setlength{\parsep}{0pt}
\setlength{\topsep}{1pt}
    \item \textsc{MI}: the best PACIFIER variant scores 0.264, compared with
    0.261 for Greedy and 0.262 for ExtremeExpressed-FI, leaving only a
    \textbf{1.06\%} gap to the best non-PACIFIER method.
    \item \textsc{ME}: PACIFIER-RL-FI scores 0.069, only \textbf{2.61\%}
    above Greedy, while outperforming ExtremeNeighbours-FI by \textbf{44.50\%}.
    \item \textit{node removal}: PACIFIER-RL scores 0.23, compared with
    0.285 for Greedy, giving a \textbf{19.32\%} improvement.
    \item \rev{Feedback-free vs. FI PACIFIER}: averaged across all three tasks, standard
    PACIFIER is only \textbf{0.21\%} below its full-information counterpart.
    In \textsc{MI} and \textit{node removal}, the best \rev{feedback-free} variant is
    even \textbf{1.19\%} and \textbf{1.16\%} better than its FI counterpart.
    In \textsc{ME}, \rev{feedback-free} PACIFIER-RL is only \textbf{2.96\%} below
    PACIFIER-RL-FI.
    \item GNN-GreedyExt is consistently weaker.
\end{itemize}
\vspace{-0.45em}

\vspace{-0.25em}
\subsection{Runtime: \rev{Feedback-Free} and Full-information Settings}
\label{subsec:runtime_both}

Runtime results give the efficiency side of the \rev{feedback-free} argument. \rev{Feedback-free}
PACIFIER avoids repeated equilibrium recomputation while preserving nearly the
same ANP quality as FI PACIFIER.

\begin{figure}[!t]
    \centering

    \begin{subfigure}[t]{\linewidth}
        \centering
        \includegraphics[width=0.98\linewidth]{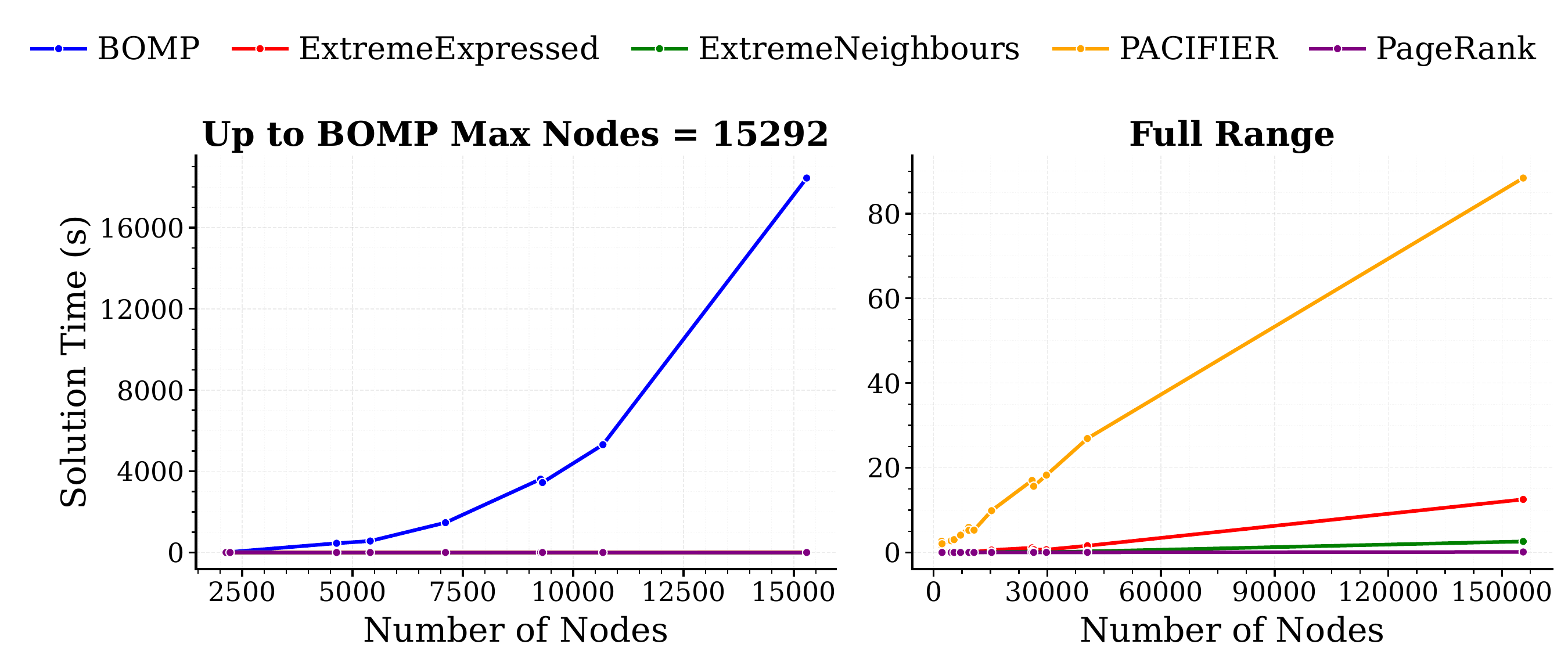}
        \caption{\rev{Feedback-free runtime} on real-world graphs.}
        \label{fig:runtime_real}
        \label{fig:runtime_with_bomp}
        \label{fig:runtime_without_bomp}
    \end{subfigure}

    \vspace{-0.2em}

    \begin{subfigure}[t]{\linewidth}
        \centering
        \includegraphics[width=0.98\linewidth]{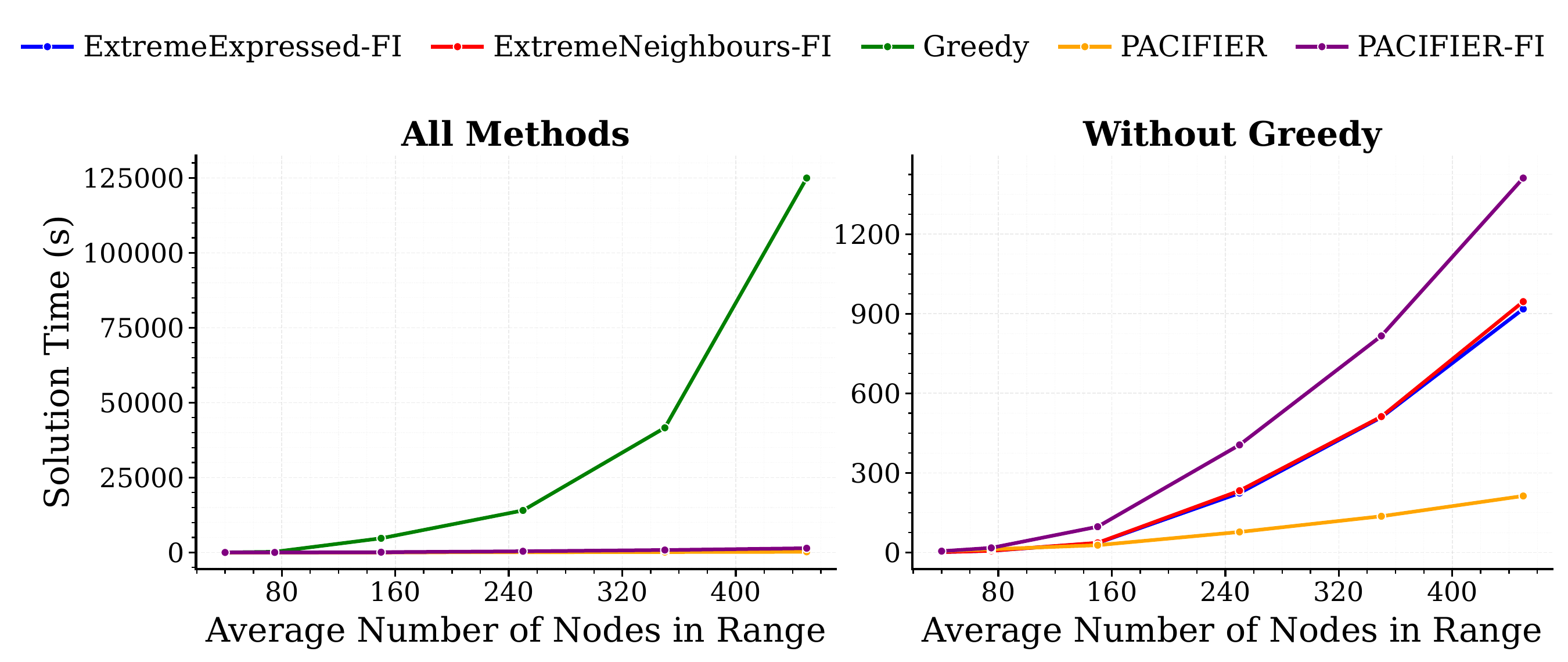}
        \caption{Full-information runtime on synthetic graphs.}
        \label{fig:runtime_fullinfo}
        \label{fig:runtime_fullinfo_all}
        \label{fig:runtime_fullinfo_no_greedy}
    \end{subfigure}

    \vspace{-0.2em}
    \caption{Runtime comparison.}
    \label{fig:runtime_combined}

    \vspace{-0.55em}
    \begin{minipage}{\linewidth}
        \centering
        \makeatletter
        \def\@captype{table}
        \makeatother
        \caption{Average ANP performance-loss percentage on real-world datasets.}
        \label{tab:ablation_summary}
        \setlength{\tabcolsep}{4.25pt}
        \renewcommand{\arraystretch}{1.00}
        \begin{tabular}{lccc}
            \toprule
            Setting & without\_aux & without\_anchor & without\_mark \\
            \midrule
            \textsc{MI} + Greedy & $-0.62\%$ & $+1.45\%$ & $-29.05\%$ \\
            \textsc{MI} + RL     & $-5.38\%$ & $-3.30\%$ & $-4.36\%$ \\
            \textsc{ME} + Greedy & $-0.93\%$ & $-8.19\%$ & $-36.15\%$ \\
            \textsc{ME} + RL     & $-13.85\%$ & $-21.83\%$ & $-28.68\%$ \\
            \bottomrule
        \end{tabular}
    \end{minipage}
\end{figure}

\vspace{-0.5em}
\paragraph{\rev{Feedback-free runtime} on real-world graphs.}
Figure~\ref{fig:runtime_real} shows the \rev{feedback-free runtime} on real-world graphs.
PACIFIER uses feed-forward inference: a $K$-layer GraphSAGE encoder costs
$O(K(n+m))$, decoder scoring is parallel over nodes, and ranking adds
$O(n\log n)$.

\begin{itemize}
\setlength{\itemsep}{0pt}
\setlength{\parskip}{0pt}
\setlength{\parsep}{0pt}
\setlength{\topsep}{1pt}
    \item BOMP grows sharply and reaches about $1.8\times 10^4$ s at the
    largest supported graph.
    \item After excluding BOMP, PACIFIER reaches about 89 s on the largest
    graph, while the next slowest baseline is around 13 s.
\item PACIFIER is slower than light heuristics but scales well and avoids
BOMP's prohibitive dense influence-matrix operations.
\end{itemize}
\vspace{-0.45em}

\vspace{-0.5em}
\paragraph{Full-information runtime on synthetic graphs.}
As shown in Fig.~\ref{fig:runtime_fullinfo}, we also measure runtime when
methods explicitly recompute intermediate
settled-opinion states during sequential selection.

\begin{itemize}
\setlength{\itemsep}{0pt}
\setlength{\parskip}{0pt}
\setlength{\parsep}{0pt}
\setlength{\topsep}{1pt}
    \item Greedy is the dominant bottleneck, reaching about $1.25\times 10^5$ s
    in the 400--500 node range.
    \item In the same range, standard PACIFIER is around 210 s, while
    ExtremeExpressed-FI, ExtremeNeighbours-FI, and PACIFIER-FI are around 920 s,
    950 s, and 1410 s, respectively.
    \item Full-information feedback is useful as an oracle comparison, but it
    introduces substantial overhead. \rev{Feedback-free} PACIFIER gives the better
    deployment-time trade-off.
\end{itemize}
\vspace{-0.45em}

\vspace{-0.6em}
\subsection{Ablation Study}
\label{subsec:ablation}

We ablate three representation inputs: polarization-related auxiliary features,
the initial-opinion anchor, and the intervention-history mark. For each of
\textsc{MI} and \textsc{ME}, we evaluate both PACIFIER-Greedy and PACIFIER-RL.

Table~\ref{tab:ablation_summary} shows that history-aware representation is
necessary. The intervention-history mark is most critical, with removals
degrading greedy \textsc{MI} and \textsc{ME} by \textbf{29.05\%} and
\textbf{36.15\%}; under RL, all components matter, especially in \textsc{ME},
where removing auxiliary features, the anchor, and the mark degrades
performance by \textbf{13.85\%}, \textbf{21.83\%}, and \textbf{28.68\%}.

\vspace{-0.7em}
\section{Conclusion}
\label{sec:conclusion}

We presented \textbf{PACIFIER}, a unified graph-learning framework with an RL
variant for FJ-based opinion polarization moderation. PACIFIER preserves the
canonical \textsc{ModerateInternal} and \textsc{ModerateExpressed} semantics,
but reformulates them as \rev{feedback-free autoregressive sequential planning}
evaluated by Accumulated Normalized Polarization (ANP). To address state
aliasing in fixed-topology continuous-state moderation, PACIFIER uses
history-aware node features, initial-opinion anchors, selected-node marks,
and scale-normalized auxiliary signals. Empirically, PACIFIER trains only on
synthetic graphs with fewer than 50 nodes and transfers to Twitter networks
with up to 155{,}599 nodes, improving \textsc{ME}, continuous-\textsc{ME},
cost-\textsc{ME}, and \textit{node removal}, while staying within 0.21\% of
its full-information counterpart.
\rev{Future work will study PACIFIER under noisy, missing, and partially observed
graph--opinion information.}

\begin{acks}
\begingroup
\setlength{\emergencystretch}{3em}
The authors acknowledge the financial support from the National Natural Science
Foundation of China (Grant Nos. 62476173, 62276171, 62002233, 61972145,
62532007), CCF-Huawei Populus Grove Fund (Grant Nos. CCF-HuaweiFM2024004),
the Shenzhen Fundamental Research-General Project (Grant Nos.\allowbreak\ 
JCYJ20240813142610014,\allowbreak\ JCYJ20240813141503005,\allowbreak\ 
ZDCY20250901110940006), Guangdong Basic and Applied Basic Research Foundation
(Grant Nos. 2024A1515 011938 and 2020B1515120028), Major Special Project for
Philosophy and Social Sciences Research of the Ministry of Education
(Grant No. 2025JZDZ010).
\par
\endgroup
\end{acks}

\section*{GenAI Usage Disclosure}
The authors used ChatGPT to assist with language polishing, LaTeX formatting,
manuscript editing suggestions, and code-debugging discussions. All scientific
claims, experimental design, data analysis, code, and final manuscript content
were verified by the authors, who take full responsibility for the work.

\bibliographystyle{ACM-Reference-Format}
\balance
\bibliography{references}

\end{document}